\documentclass[
twocolumn,showpacs,superscriptaddress, preprintnumbers , amsmath,amssymb,apssamp, letterpaper,  prl]{revtex4-1}

\usepackage[utf8]{inputenc}
\usepackage{graphicx}% Include figure files
\usepackage{dcolumn}% Align table columns on decimal point
\usepackage{bm}% bold math
\usepackage{amsmath}
\usepackage{multirow}

\usepackage[colorlinks=true]{hyperref}
\usepackage{amsmath}
\usepackage{amssymb}
\usepackage{array,booktabs}
\usepackage{graphicx}
\usepackage{amsbsy}

\begin{document}
\title{Design and Characteristics of a Population Inversion X-ray Laser Oscillator}
\author{A.	Halavanau}
\affiliation{SLAC National Accelerator Laboratory, Menlo Park, CA 94025, USA}
\author{A. Benediktovitch}
\affiliation{Center for Free Electron Laser Science, DESY, Hamburg 22761, Germany}

\author{A.A. Lutman}
\affiliation{SLAC National Accelerator Laboratory, Menlo Park, CA 94025, USA}

\author{D. DePonte}
\affiliation{SLAC National Accelerator Laboratory, Menlo Park, CA 94025, USA}

\author{D. Cocco}
\affiliation{Lawrence Berkeley National Laboratory, Berkeley, CA 94720, USA}

\author{N. Rohringer}
\affiliation{Center for Free Electron Laser Science, DESY, Hamburg, 22607, Germany}
\affiliation{Department of Physics, Universit{\"a}t Hamburg, Hamburg 22761, Germany}

\author{U. Bergmann}
\affiliation{SLAC National Accelerator Laboratory, Menlo Park, CA 94025, USA}

\author{C. Pellegrini}
\affiliation{SLAC National Accelerator Laboratory, Menlo Park, CA 94025, USA}

\begin{abstract}
    Oscillators are at the heart of optical lasers, providing stable transform limited pulses. In contrast, X-ray free electron lasers use self-amplified spontaneous emission (SASE), resulting in large stochastic intensity and spectral fluctuations. Amplified spontaneous emission (ASE) of the $K\alpha_1$ line has been recently observed for Ne gas, Cu compounds and Mn solutions at the LCLS and SACLA X-ray free electron lasers (XFELs), using an X-ray SASE pulse as a pump to create population inversion. Here we describe the physics and realization of an X-ray laser oscillator (XLO) based on periodically pumping a Cu compound gain medium in a tunable Bragg cavity with a SASE pulse train, generating intense ($\sim$ 5 x 10$^{10}$ ph/pulse), fully coherent, transform limited 8 keV pulses with 48 meV spectral resolution.
We also discuss extending these results to other elements to operate XLO from about 5 to 12 keV,
improving X-ray-based research beyond current capabilities.

\end{abstract}

%\begin{abstract}
%Amplified spontaneous emission (ASE) of the $K{\alpha}_1$ line and seeded amplified emission of the $K\beta$ lines have been recently observed for Ne gas, Cu compounds and Mn solutions at the LCLS and SACLA X-ray free electron lasers (XFELs), using an X-ray SASE pulse as a pump to create population inversion. In this paper, we propose to use a train of SASE pump pulses, e.g. from LCLS, impinging on a liquid jet of cupric nitrate in a tunable crystal Bragg cavity, to realize an X-ray laser oscillator (XLO) at the Cu $K{\alpha}_1$ line. XLO is a novel X-ray source generating fully coherent, transform limited 8 keV X-ray pulses with high photon numbers (5$\times 10^{10}$ ph/pulse) and high spectral resolution (48 meV), tunable over a $\sim$3 eV range. Using different elements, e.g. Mn, Fe, Zn, etc., XLOs with discrete photon energies in the range of 5 to 12 keV are feasible. With its unique characteristics XLOs will extend research in atomic and molecular science beyond current capabilities.
%\end{abstract}


\maketitle

%\section{Introduction and XLO description}
Since the time Maiman built the first ruby laser in 1960, developing an X-ray laser has been a major goal in laser physics.
%A laser generating high intensity, coherent X-ray pulses at Angstrom wavelength and femtosecond pulse duration - the characteristics time and space scale for atomic and molecular phenomena - allows imaging of periodic and non-periodic systems, non-crystalline states, studies of dynamical processes in systems far from equilibrium, nonlinear science, opening a new window an atomic and molecular phenomena of interest to biology, chemistry materials science and physics \cite{10.1039/9781782624097}.
Several successful attempts were made to create a population inversion X-ray laser, however with limited peak power and tunability. It included the use of strong optical lasers to pump plasmas, and even nuclear weapons, e.g. in Excalibur/Dauphin experiment \cite{Hecht:08, ChaplineWood, PhysRevLett.54.110, PhysRevLett.55.1753, Suckewer_2009}.
X-ray free-electron lasers (XFELs), first proposed in 1992 and developed from the late 1990s to today \cite{PELLEGRINI1993223, RevModPhys.88.015006}, are a revolutionary new tool to explore matter at the atomic length and time scale, with high peak power, transverse coherence, femtosecond (fs) pulse duration, nanometer to angstrom wavelength range \cite{RevModPhys.88.015007, 10.1039/9781782624097}.  They have also been shown to be a very effective pump for creating population inversion in an atomic gain medium, as has been demonstrated in experiments demonstrating amplified spontaneous emission (ASE) in a Ne gas \cite{Rohringer2012}, Cu metal foils \cite{Yoneda} and recently Mn solutions \cite{PhysRevLett.120.133203}. An intense pump pulse creates a 1s core-hole and spontaneously emitted $K{\alpha_1}$ fluorescence photons along the direction of the population inversion are amplified. Large values of gain (up to $2\times10^6$) were observed and $K{\alpha_1}$ ASE pulses of up to $8\times10^7$ photons/pulse are generated starting from noise \cite{PhysRevLett.120.133203, Kroll:16}.
\begin{figure}
    \centering
    \includegraphics[width=1\linewidth]{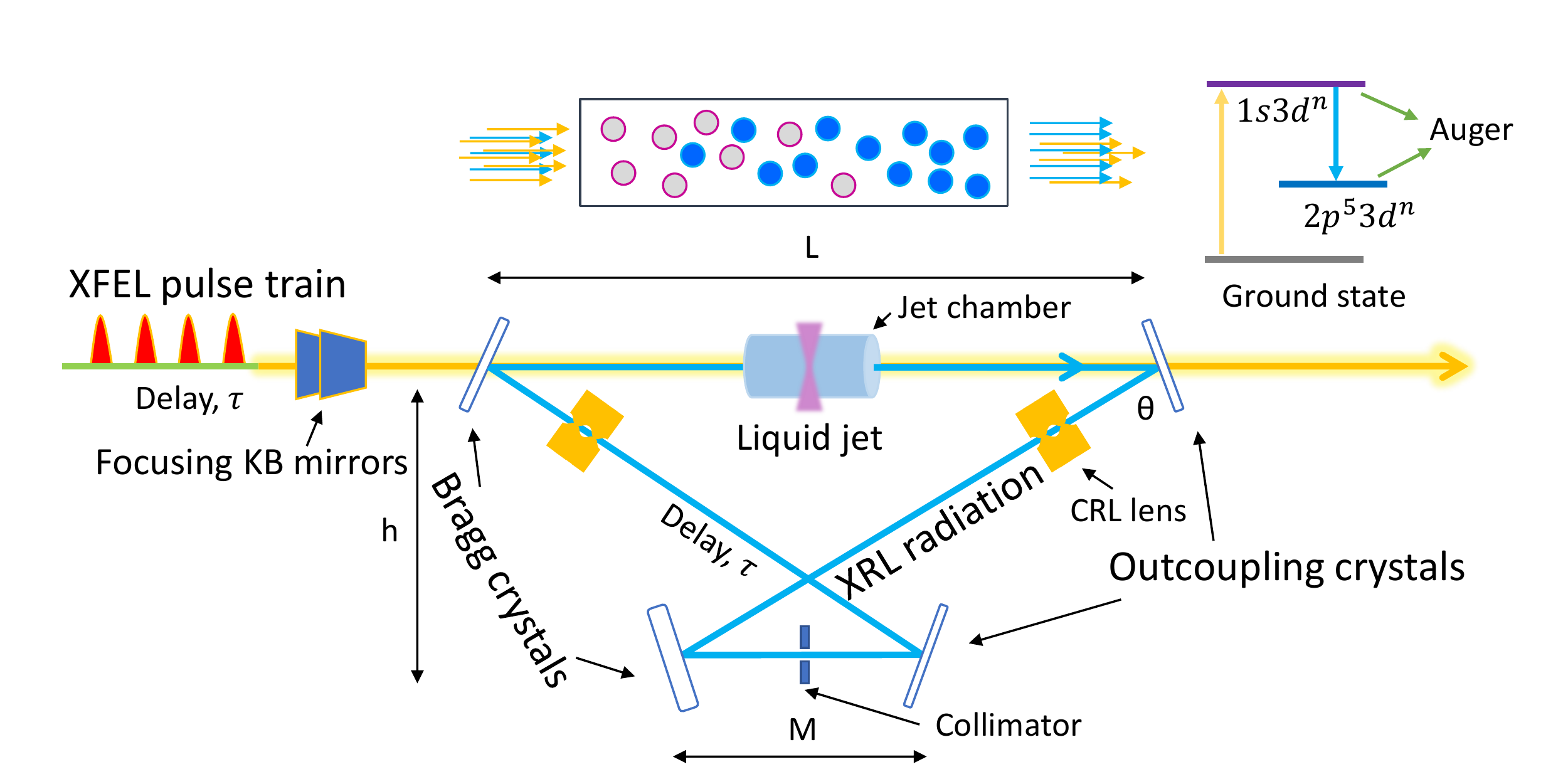}
    \caption{XLO schematics. A train of SASE pulses with spacing $\tau$ impinges on a liquid jet, generating population inversion and photon emission. The amplified $K{\alpha_1}$ signal is recirculated with a roundtrip time that matches $\tau$ and outcoupled after $N$ passes. In the figure, the separation between X-ray pump pulses and the cavity size is not to scale.  }
    \label{xlo:fig1}
\end{figure}

Most XFELs operate also in ASE, or self-amplified spontaneous emission (SASE) as it is called in the XFEL community. In SASE mode \cite{Bonifacio:1984qs, DERBENEV1982415}, the X-ray pulse starts from electron shot-noise.
%Due to the stochastic nature of this process, SASE X-ray pulses have limited longitudinal coherence and significant shot-to-shot fluctuations. %The same is true for ASE in atomic systems with population inversion.
To improve longitudinal coherence and fluctuations,
seeding with external lasers has been successfully
implemented for wavelengths larger than a few nanometers
\cite{PhysRevA.44.5178, Allaria2012, Allaria2013, Stupakov2009, Ribic2019}.
For harder X-rays, self-seeding systems have been employed to reduce the bandwidth by an order of magnitude w.r.t.  SASE  pulses,  e.g. at LCLS from about $10^{-3}$ to $10^{-4}$, but the pulses are not transform limited, require further monochromatization and suffer from large intensity fluctuations \cite{PhysRevLett.114.054801, doi:10.1080/09500340.2011.586473, Amann, PhysRevLett.113.254801, Osaka:rx5047, Inoue2019, PhysRevAccelBeams.22.080702, Colocho2013INCREASEDSR}. 
To produce transform limited FEL pulses, one can reduce the pulse duration down to a single coherent SASE spike, making the electron bunch length shorter than the cooperation length \cite{REICHE200845}. The  use of emittance  spoilers \cite{PhysRevLett.92.074801, doi:10.1063/1.4935429, doi:10.1063/1.4990716}, the combination of fresh-slice, cascaded amplification \cite{Lutman2016, Lutman:2018cot}, non-linear compression schemes \cite{PhysRevLett.119.154801} and recently eSASE \cite{JoeNature} have succeeded in producing such pulses.  However, in these schemes the coherence   is achieved by decreasing the pulse duration, resulting in a larger bandwidth and lower photon flux compared to self-seeded modes as well as large shot-to-shot fluctuation. To drastically increase the photon flux, a double bunch FEL (DBFEL)  scheme at LCLS has been investigated \cite{PhysRevAccelBeams.20.030701, doi:10.1063/1.4980092, Halavanau:co5121}. To generate fully coherent hard X-ray pulses, an X-ray FEL oscillator (XFELO) \cite{PhysRevLett.100.244802, COLELLA198441, PhysRevLett.96.144801, Adams:2019njp} has been proposed. Since the lasing medium is an electron bunch in a long, typically 100 m, undulator, XFELO requires a total of few hundred meters long diamond Bragg crystal cavity wrapped around the undulator. Mechanical stability, radiation focusing, temporal-transverse beam overlap in the cavity, and crystal alignment angular tolerances, of the order of 10 nrad, are important technical issues. Ultimately, XFELO could operate with a high-repetition rate temporally equispaced set of long electron bunches, achieving ultra-narrow bandwidth, about $10^{-7}$, and very large average brightness at hard X-rays.

%In this Letter, we propose a novel approach for producing stable, high power, narrow-band hard X-ray pulses based on employing a population inversion gain medium within an oscillator cavity that is periodically pumped by a train of SASE pulses from an XFEL, as schematically shown in Fig. \ref{xlo:fig1}. $Amplified Spontaneous Emission (ASE) has been observed and studied for Ne gas \cite{Rohringer2012}, Cu metal foils \cite{Yoneda} and recently for Mn solutions \cite{PhysRevLett.120.133203}. Here, an intense pump pulse creates a $1s$ core-hole and spontaneously emitted $K{\alpha}$ fluorescence photons along the direction of the population inversion are amplified. Depending on the pump intensity and gain medium, very high values of gain (up to $2 \times 10^6$) were observed and $K{\alpha_1}$ ASE pulses of up to $8 \times 10^7$ photons/pulse were created starting from noise \cite{PhysRevLett.120.133203}. 

The XLO system described here and shown in Fig. \ref{xlo:fig1} starts with a SASE pump pulse that creates a 1s corehole population inversion followed by ASE in a liquid jet. We consider, as an example, a jet of concentrated cupric nitrate (Cu(NO$_3$)$_2$) solution, with about 4.2 Cu atoms per nm$^3$. After the ASE gain in the first pass, the $K{\alpha_1}$ signal is recirculated through the cavity, including losses, and matched with the arrival of the subsequent XFEL pulses, creating seeded stimulated emission. Once saturation is achieved after 4-8 round trips, the pulse is outcoupled from the cavity. We have chosen a cupric nitrate solution as a gain medium because the copper $K{\alpha_1}$ photon energy at 8 keV is widely used in X-ray science. Furthermore, cupric nitrate is soluble in high concentrations, does not contain any heavy elements that would reduce the gain by absorbing 8 keV photons, and is straightforward to handle. While we focus on the Cu $K{\alpha_1}$ line and a Si (444) crystal Bragg cavity geometry that matches its wavelength, the XLO concept is not limited to Cu $K{\alpha_1}$ line, and other choices of the lasing medium and matching cavity configurations are possible. In fact, switching between different lasing media requires a minimal change in the cavity geometry. For example, the Fe $K{\alpha_1}$ line at 6.4 keV using a Si (333) Bragg reflection with 67.9 degree Bragg angle, can be easily realigned to match the Zn $K{\alpha_1}$ line at 8.6 keV switching to the Si (444) reflection with a slight realignment to a 66.3 degree Bragg angle.

The XLO Bragg cavity length $C_l$ can range from approximately 50 cm to a few meters, depending on the value of SASE train separation $\tau$ from 1.7 - 10 ns.
The cavity length $C_l$ and lateral size $h$ are related to the upper and lower crystals separation $L$, $M$ and Bragg angle $\theta$ by: 
%$C_l = (L+M)(1-1/\cos{2\theta})$ and %$h=-\frac{1}{2}(L+M)\tan{2\theta}$.
$(L+M) = C_l / (1-1/\cos{2\theta})$ and $h=-\frac{1}{2}(L+M)\tan{2\theta}$, where $C_l = c \tau$.
The optimal size is a compromise between ensuring sufficient time to replace the lasing medium between subsequent pulses in the train, and guaranteeing sufficient mechanical/temperature stability of the cavity. The resulting X-ray laser pulses are highly monochromatic (as defined by the Bragg optics in the cavity), transform limited, and provide one order of magnitude higher pulse energy with significantly enhanced stability compared to monochromatized SASE pulses. While the tunability is limited to the width of the emission line (3-5 eV), such XLOs can be built for various discrete energies using different emission lines that match the desired experimental parameters. The unprecedented pulse quality resulting from this approach will improve many X-ray experiments \cite{Adams:2019njp}. 

KB mirrors, upstream of the XLO, can be used
to effectively pump the lasing medium, as previously employed in the LCLS Mn ASE experiments \cite{PhysRevLett.120.133203}. Such a setup exists at the LCLS Coherent X-ray Imaging (CXI) instrument, and similar systems are available at the SACLA BL3 and Eu-XFEL SFX beamlines. 
For the wavelength considered in this study, Silicon and Diamond are optimal choices for the cavity optics. The characteristics of these two crystals for various diffraction planes, matching the Cu $K{\alpha_1}$ emission line are given in Tab. \ref{table-crystals} and calculated with dynamical diffraction code XOP \cite{XOPcode}.
The ASE beam divergence depends on the aspect ratio of beam size and gain length and is of order 1 mrad, well above the Si crystal angular acceptance for good reflectivity. Hence, on the first pass the cavity losses are large, and need to be overcompensated by a large XLO laser gain ($10^{4}$). At the following passes the losses will be much smaller since the angular spread of the amplified photons matches the crystal acceptance.
%\section{Theoretical model}
%\label{section:xlo_theory}
To analyze the XLO performance we first consider the amplification process in the lasing medium. The emission from the population-inverted medium is calculated using a correlation function approach \cite{UweKB, PhysRevA.99.013839, PhysRevLett.123.023201}. We use the {\sc XATOM} software \cite{Jurek:zd5003, PhysRevA.83.033402, RevModPhys.39.125, GLATZEL200565} to evaluate the photoionization cross section of the copper atom 1s level and the Auger decay time to be 0.0324 Mb and 823 as. 
The evolution of atomic populations is computed, in a 1D model approximation, via a two-point correlation function of the atomic coherences (polarization) and two-time correlation function of fields. 

%\begin{figure}
 %   \centering
 %       \includegraphics[width=0.32\linewidth]{Cu_levels.pdf}
%    \includegraphics[width=0.54\linewidth]{pcsCu.pdf}\\
  %  \includegraphics[width=0.85\linewidth]{Natoms_new.pdf}
   % \caption{
   % Atomic transitions schematics and photoionization cross section of Cu $K{\alpha_1}$ \cite{RevModPhys.39.125, GLATZEL200565} (top row). 1 Mbarn is $10^{-4}$ nm$^2$. 
    %Number of Cu atoms in Cupric Nitrate, Cupric Sulphate solutions and pure copper pumped volume as a function of the SASE pulse radius and corresponding power density (bottom row). 
    %}
%    \label{xlo:pcs}
%\end{figure}
At the initial stage of the emission, the spontaneously emitted radiation (described in terms of field correlation function) triggers the amplification process. During the propagation through the population inverted medium, the radiation is amplified and the contribution of the spontaneous emission becomes negligible compared to amplified stimulated emission. At this point, the field correlation function can be factorized into classical field amplitudes. Corresponding equations (after neglecting the spontaneous emission contributions) can be factorized as well and become equivalent to the Maxwell-Bloch equations. The field obtained is propagated through the Bragg crystal cavity elements and is used as a seed for the next round of amplification. Starting with the second round, the spontaneous $K{\alpha_1}$ emission along the gain medium is much weaker than the recirculated seeding field and the contribution of the spontaneous emission terms can be neglected leading to a description in terms of Maxwell-Bloch equations. In a similar way, Maxwell-Bloch equations are used for the subsequent rounds of amplification. 

\begin{figure}
    \includegraphics[width=0.47\linewidth]{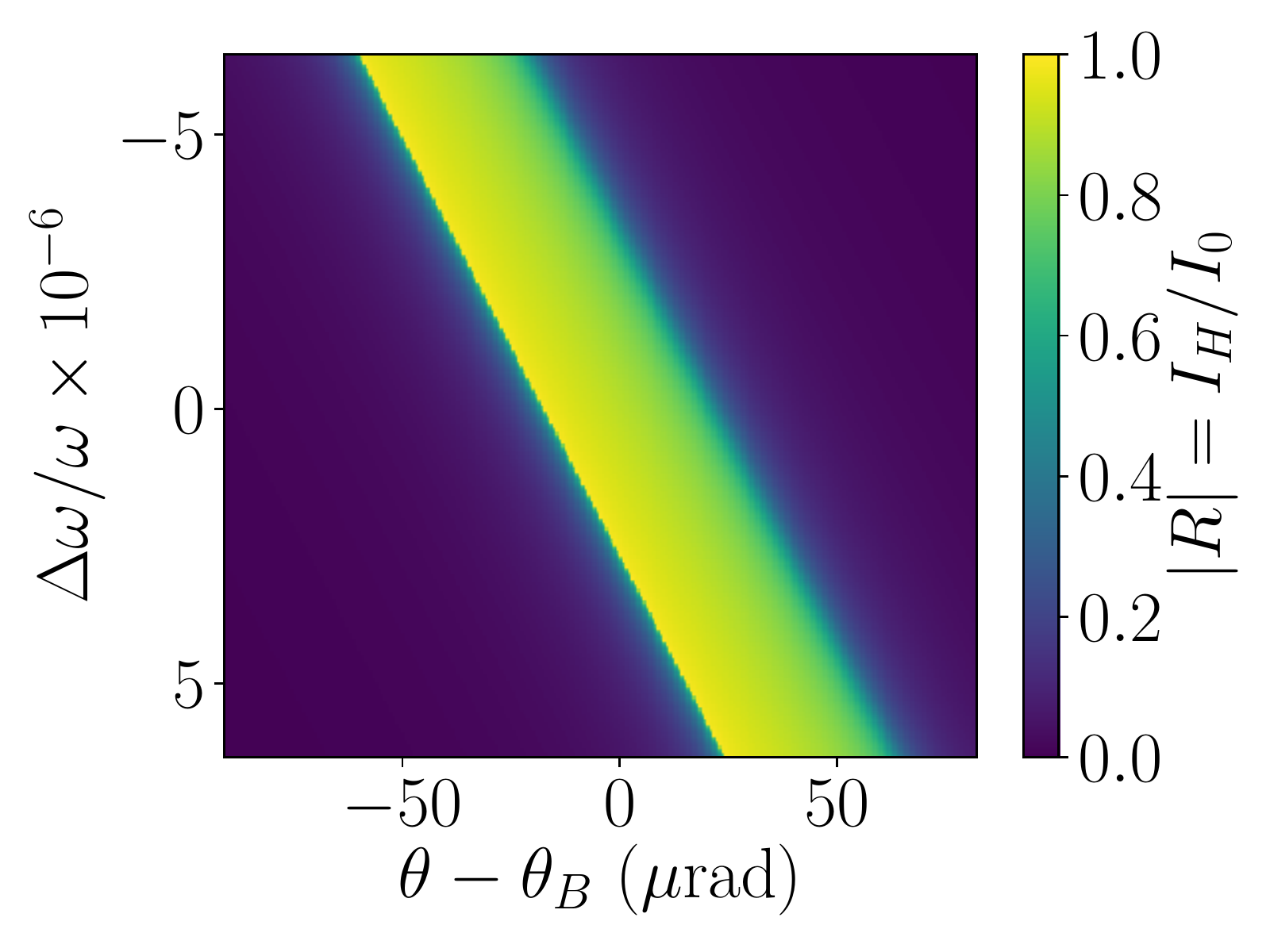}
    \includegraphics[width=0.471\linewidth]{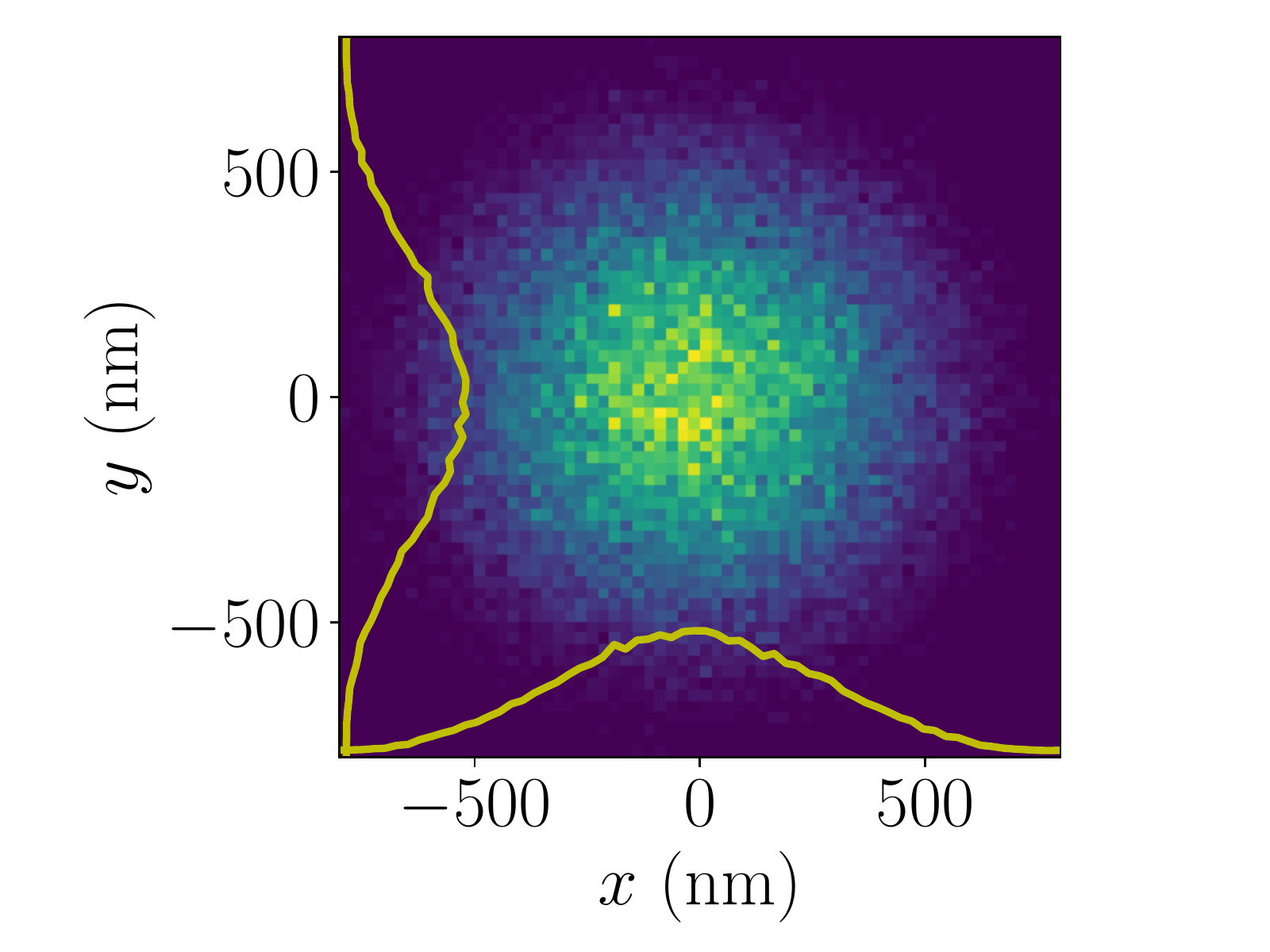}
    \includegraphics[width=0.5\linewidth]{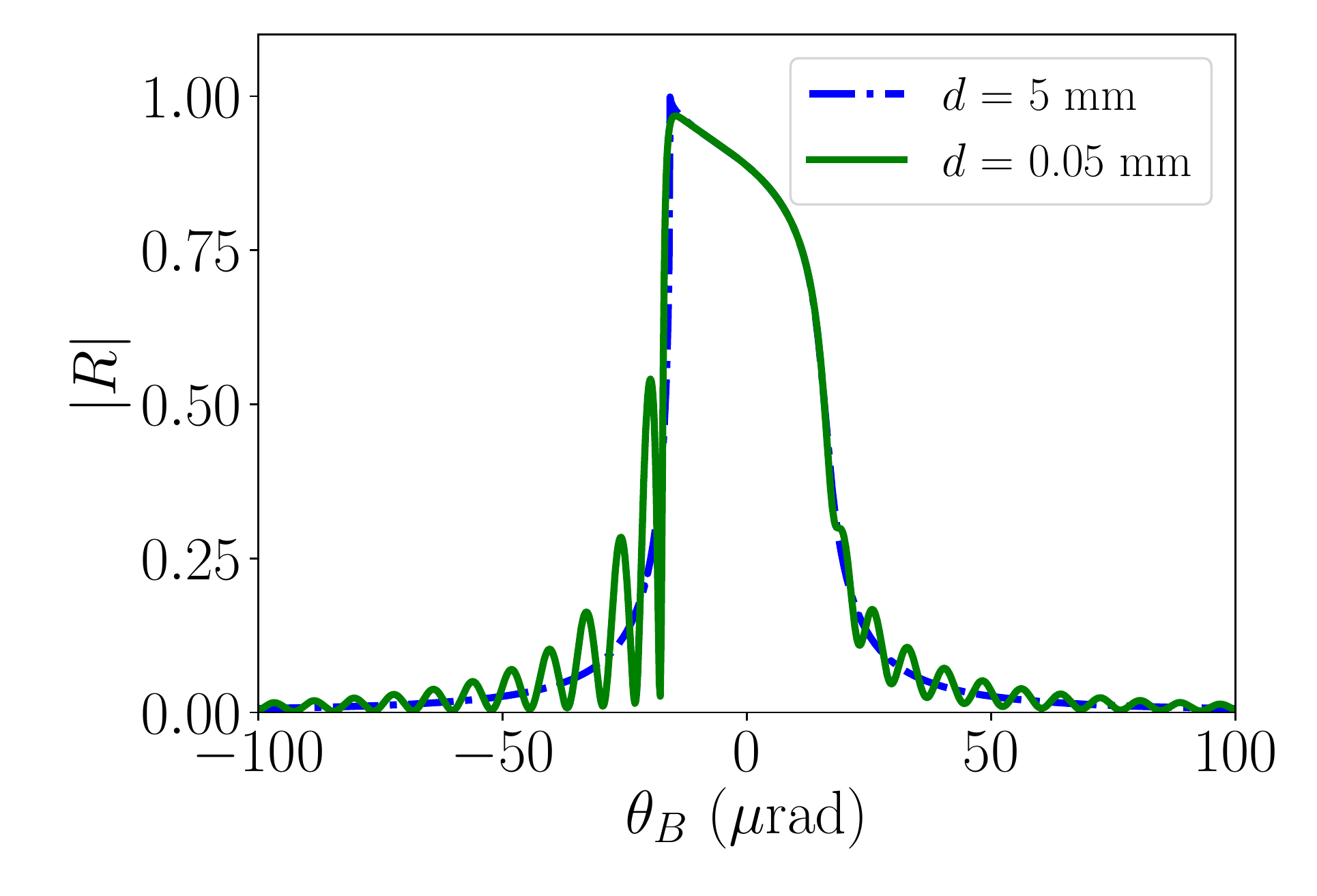}
      \caption{DuMond diagram of Si (444) crystal at 8048 eV (top left). Refocused ASE radiation at the jet location after two sets of Be lenses with state of the art shape profiles (top right).
      Darwin curve for Bragg reflection of 8048 eV photons, corrected for refraction for the cases of thick ($d$ = 5 mm) and thin ($d$ = 50 $\mu$m) flat Si (444) crystals at room temperature (bottom).}
    \label{xlo:darwin}
\end{figure}

%\section{XLO cavity design}
%\label{section:xlo_cavity}
Photons are recirculated using a tunable four bounce bowtie $(+--+)$ Bragg crystal cavity \cite{doi:10.1063/1.1651874} , as shown in Fig. \ref{xlo:fig1}. Its optical properties are graphed in Fig. \ref{xlo:darwin}.
The bowtie cavity is tuned to ensure the exact spectral alignment of the Cu $K{\alpha_1}$ peak wavelength with the Bragg reflection angle and temporal alignment with the pulse train separation $\tau$. The first crystal (upper left) is thin and highly transparent for the incident pump pulse at 9 keV, yet highly reflective at the Bragg angle of the Cu $K{\alpha_1}$ peak. One of the other three crystals (e.g. upper right in Fig. \ref{xlo:fig1}) is also thin but highly reflective at the chosen wavelength. It is used to outcouple the X-ray pulse when it has reached saturation after several amplification cycles. Outcoupling is done by switching to a status with large transmission and near zero reflection.  Several techniques for radiation outcoupling, e.g. a small fast rotation or heating with a laser pulse, can be employed and are currently being studied \cite{Kolodziej:te5013, Shvydko:2019gvv, Freund_2019}. 
We choose the backscattering Bragg geometry as it provides a large angular acceptance while reducing the bandwidth, see Tab.\ref{table-crystals} and Fig. \ref{xlo:darwin}. The Darwin width of the Bragg reflection defines the spectral-angular acceptance and the corresponding X-ray pulse length and lasing medium radius. Using the Si (444) reflection gives a 48 meV bandwidth, while the dimanond C*(400) reflection gives a 65 meV bandwidth with a more then two times smaller angular acceptance.The cavity tolerances in the crystal angle and position have been estimated to be of the order of 1 $\mu$m and 1 $\mu$rad, well within the capability of present day crystal support and alignment systems.

\begin{table}
\centering
  \begin{tabular}{c c c c c c }\hline
\hline
%\multicolumn{1}{|c|}{\multirow{2}{*}
(hkl) &	$\theta_B$ (deg) &	$\Delta \theta$  ($\mu$rad)& $\Delta \omega$ (meV) & $\Delta \omega / \omega \cdot 10^{-6}$ & $\tau$(fs)    \\
\hline
Silicon &&&&&\\
%111	& 14.2  & 34.7	& 1103.7	& 137.1	& 1.6 \\
%220	& 23.7	& 26.5	& 485.9	& 60.4	& 3.7 \\
%311	& 28.1	& 15.5	& 233.6	& 29.0	& 7.7 \\
333	& 47.5	& 10.3	& 76.0	& 9.4	& 23.7 \\
533	& 68.5	& 12.5	& 39.6	& 5.0	& 45.5 \\
444	& 79.3	& 31.6	& 48.1	& 6.0	& 37.4 \\
\hline
Diamond &&&&&\\
%111	& 22.0	& 24.2	& 482.1	& 60.0	& 3.7\\
%220	& 37.7	& 15.9	& 168.0	& 21.0	& 10.7\\
311	& 45.7	& 9.6	& 75.4	& 9.4	& 23.9\\
400	& 59.8	& 14.0	& 65.6	& 8.1	& 27.4\\
331	& 70.3	& 13.0	& 37.4	& 4.7	& 48.1\\
\hline
\hline
\end{tabular}
\caption{\label{table-crystals}Silicon and Diamond crystal parameters for 8048 eV photons in different Bragg reflection geometries. }
\end{table}

To understand the salient features of the XLO cavity we use the formalism of Fourier optics and dynamic diffraction theory in perfect crystals \cite{RevModPhys.36.681, PhysRevSTAB.15.050706, PhysRevSTAB.15.100702, PhysRev.137.A1869}.  We select flat symmetric crystals to alleviate any additional azimuthal angle effects on the Bragg diffraction \cite{RevModPhys.36.681}. 
We evaluate the crystals Darwin reflectivity curves, thickness and Miller indices using the XOP code \cite{XOPcode}. The  curves obtained are then convoluted with the spectral content of the electromagnetic field.   
The correction to the Bragg angle $\theta_{B}$ for a Gaussian photon beam is given by a simple geometric relation $\sin{\theta_B} \approx \sin{\theta_{B_0}}(1 - \Delta \phi^2/2)$, where $\theta_{B_0}$ is the Bragg angle for a non-divergent beam  and $\phi$ is the vertical angle with respect to the plane perpendicular to the crystal surface . 
%The Bragg reflectivity map in ($\Delta\theta, \Delta\phi$) space is shown in Fig. \ref{xlo:darwin}. 
The effect of vertical divergence is of a second order for $|\Delta\phi|<10^{-3}$ rad and will leave the radiation field mostly intact in $\phi$, while selecting in $\theta$.

The cavity losses are defined by the crystal angular and energy acceptance.
%for the reflectivity to be large, nearly 0.8 to 1. 
%The ASE photons have an angular and energy spread well above those accepted by the crystals, and the corresponding loss in the initial step of the amplification process is large. On subsequent passes through the cavity the losses are defined mainly by the crystal reflectivity and CLR lenses absorption.
The ASE signal angular divergence is approximately defined by the geometry of the lasing medium 
(1 mrad vertical, 2 mrad horizontal), leading to 99.7\% cavity losses for the first pass.
The crystal reflectivity curve $|R|$ presented in Fig. \ref{xlo:darwin} (bottom), where $|R| = I_H / I_0$, for the initial radiation intensity $I_0$  and diffracted intensity $I_H$.
Furthermore, we utilize a set of two compound refractive lens (CRL) assemblies of $f=$ 52 cm and a collimator in a 4 m cavity to refocus the amplified signal on the jet after four reflections.
%to maximize the efficiency of the pump in the lasing medium.
Fourier optics simulation results corroborate this requirement; see Fig. \ref{xlo:darwin} (top right).
After the first pass, the cavity efficiency is given by the crystal reflectivity and the losses in the focusing lenses.  We estimate the total efficiency in second and later passes to be about 41\%.

Note that, based on Mn ASE results \cite{PhysRevLett.120.133203} we can obtain up to almost $10^8$ $K{\alpha_1}$ photons in the initial pass, providing more than enough signal to overcome the first pass large cavity losses. 
%\label{section:xlo_sim}
We now present the results of numerical simulations done using the one-dimensional code \cite{PhysRevA.99.013839}, for cupric nitrate as a gain medium, for different values of the pumping power and number of passes in the XLO cavity.
 The parameters for the simulations are determined by the LCLS pump power and pulse duration, by the optical cavity properties, and by the gain medium geometry.
For 9 keV LCLS SASE X-ray pulses we can assume a peak power of 30 to 50 GW with a pulse duration variable between 20 to 60 fs. Using 33.6 GW 60 fs pulse, the corresponding pulse energy is 2 mJ, well within LCLS capabilities.

The cavity crystals define a horizontal angular acceptance of 31.6 $\mu$rad; see Tab. \ref{table-crystals}. To alleviate the effect of diffraction, one must satisfy $d \Delta\theta \geq \lambda/4\pi$, which yields $d\geq 390$ nm.
The gain medium can be described as a thin cylinder of length 300 $\mu$m, defined by the jet diameter, along the pump pulse direction of propagation and an effective radius defined by the KB focusing properties for 9 keV X-ray pump pulse. In our calculation, we assume a gain medium diameter of 800 nm, 
satisfying the above condition and justifying the use of a 1D simulation code. At this spot size, the Rayleigh length is $z_R = 3.65$ mm, i.e. much longer than the jet size. 
The vertical angular acceptance is much larger than horizontal, and hence we could optimize the system by having an elliptically shaped pump pulse. While in principle, this solution can yield a better XLO performance, for sake of simplicity, we consider only the case of a gain medium with circular cross section, with the radius defined by the horizontal acceptance. In addition to varying the pump pulse intensity, the gain can also be controlled by the focusing KB optics, albeit with slightly larger cavity losses. 

\begin{figure}
    \centering
       \includegraphics[width=0.48\linewidth]{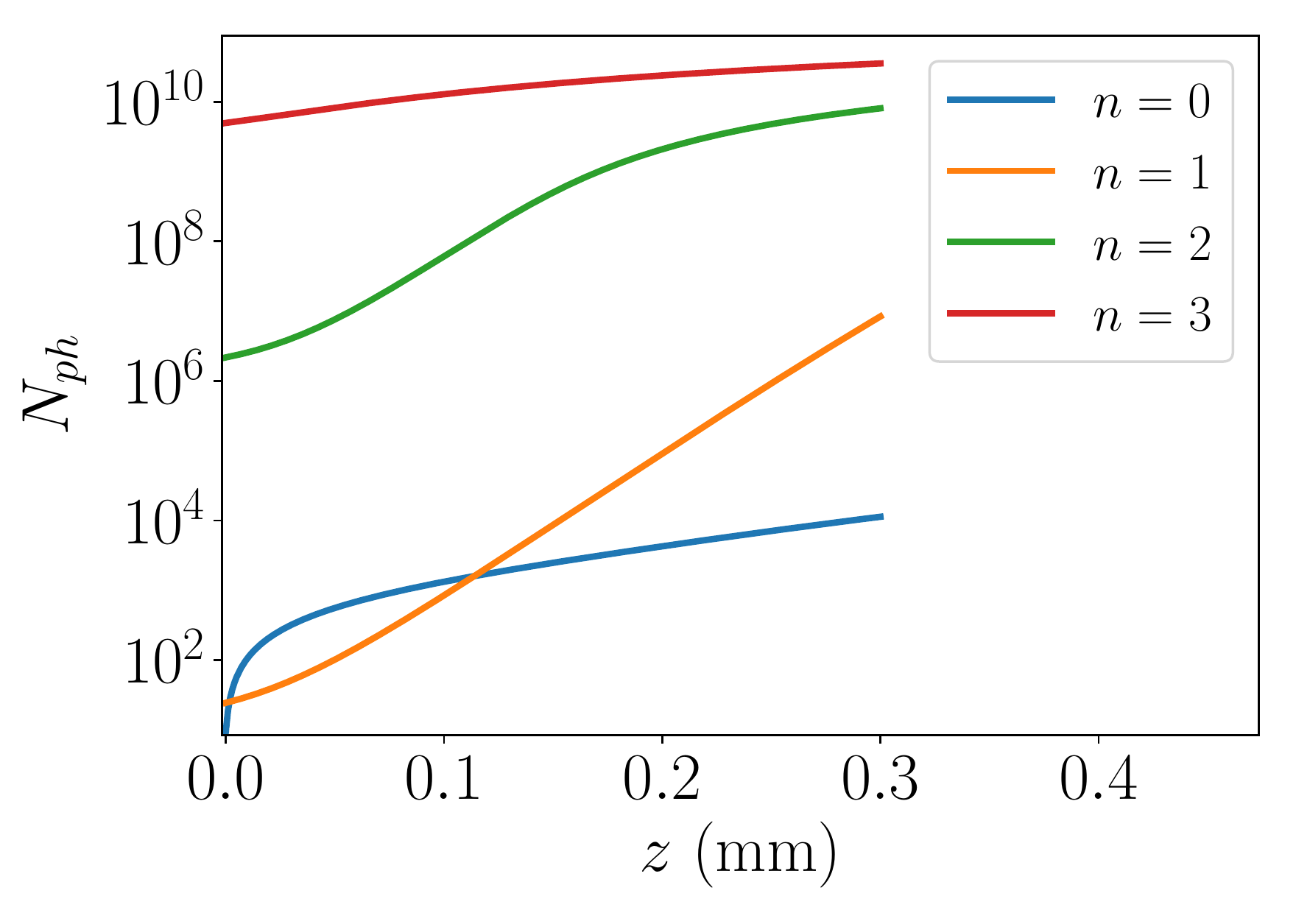}
   \includegraphics[width=0.5\linewidth]{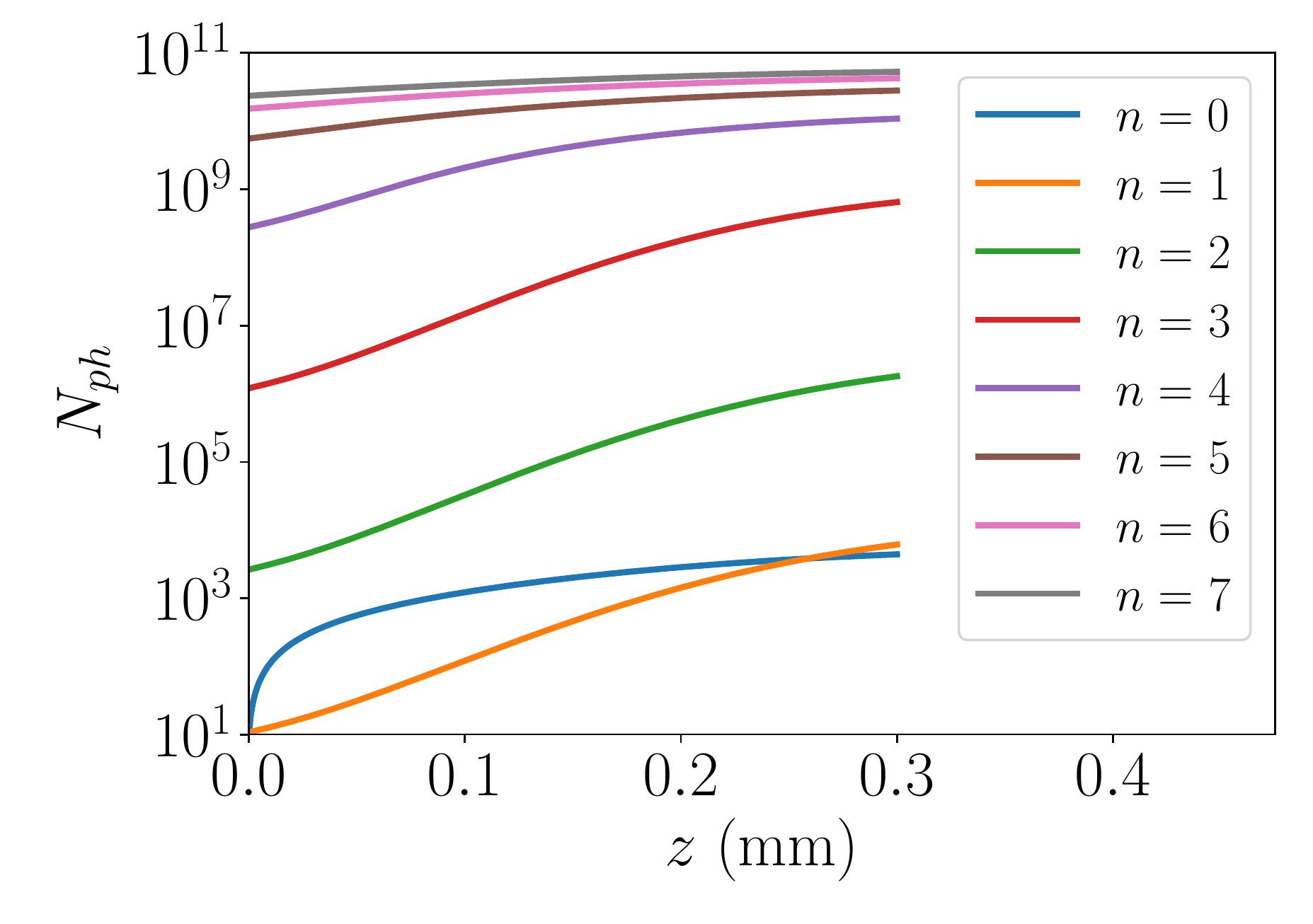}
      \includegraphics[width=0.53\linewidth]{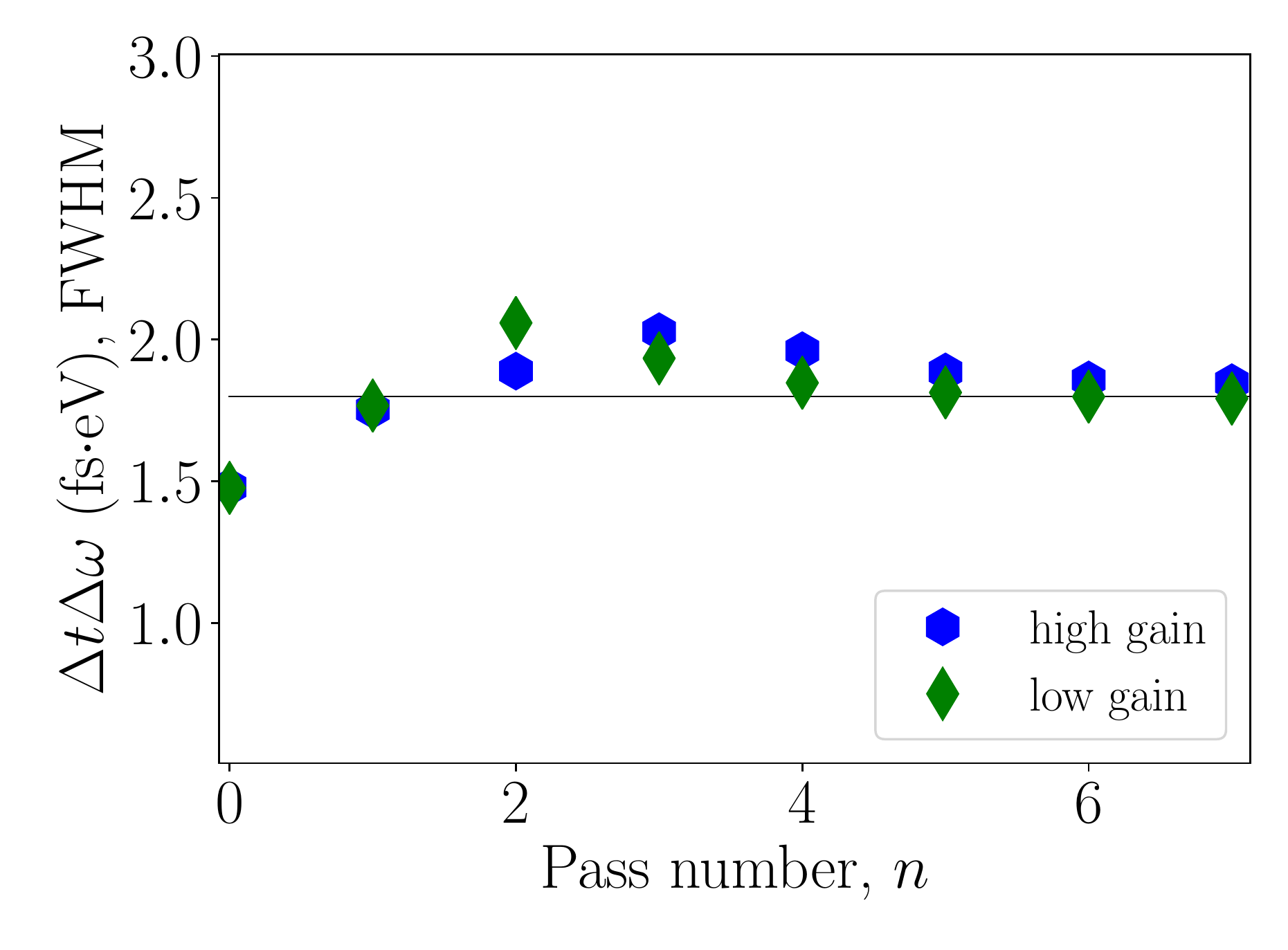}
    \caption{XLO in high-gain (top left) and low-gain (top right) regimes and resulting time-bandwidth products (bottom). The solid black line corresponds to $\Delta t \Delta\omega = 1.8$ fs$\cdot$eV. }
    \label{xlo:low}
\end{figure}

Another important quantity is the number of Cu atoms in the gain medium. Assuming 300 $\mu$m length and a 800 nm diameter, there are 6.3$\cdot 10^{11}$ Cu atoms in the lasing volume, which sets the upper limit for the number of XLO photons at saturation.
We denote the initial pass as $n = 0$, and following passes as $n = \overline{1,N}$, where $N=7$.
Depending on the pump pulse parameters, in the initial pass the ASE process generates about $10^3-10^4$ photons starting from noise. After 7 additional passes the total number of XLO photons in saturation reaches up to $7\cdot10^{10}$, as shown in Fig. \ref{xlo:low}. 

%\subsection{High- and medium- gain regimes}
We first consider high gain XLO regime: with 8 pump pulses of 2 mJ.
 In this case, rapid saturation occurs as early as in the third pass. The maximum number of photons for a 300 $\mu m$ diameter jet is about $7\cdot10^{10}$.
Thus, the XLO brightness is comparable with existing hard x-ray self-seeding (HXRSS) XFEL,
%({\color{blue} we need to discuss}), 
while offering an order of magnitude reduction in bandwidth and much better stability.
We then consider the low gain case with a SASE train of 8  pulses of 1 mJ, thus reducing the gain in the medium, and gradually building up the cavity power. 
 
In this case, radiation supermodes start to form, similarly to \cite{1071268, 1072746}. We note that the mode shape is determined by the nonlinear gain, pictured in Fig. \ref{xlo:low} and crystal reflectivity shown in Fig. \ref{xlo:darwin}. After 4 passes, the resulting XLO radiation time-bandwidth product is essentially transform-limited and is equal to 1.8 fs$\cdot$eV. The degeneracy parameter is estimated to be about $10^{10}$.

%\section{Liquid jet and LCLS CuRF linac performance}
%\label{section:xlo_jet}
\begin{table}
\centering
  \begin{tabular}{l c c }
 \hline
\hline
%\multicolumn{1}{|c|}{\multirow{2}{*}
Parameter & XLO &	XFELO \\
\hline
Gain per pass & up to 10$^6$ & 1.2 - 1.5\\
%Passes to saturation & $<$10 & $\sim$300 \\
Cavity length, m & $\sim$ 10 & $>$260 \\
Lasing medium size, m & $3\times10^{-4}$ & $\sim$100 \\
Angular tolerance, $\mu$rad & 1 & $\sim$0.01 \\ 
Number of photons (max) & $5\times 10^{10}$ & $10^{10}$  \\
Pulse length, fs & 37.4 & 530 \\
Peak power, MW & $\sim$270 & $\sim$4.7 \\
FWHM $\Delta t \Delta\omega$, fs$\cdot$eV  & 1.8 & 4.4 \\
\hline
\hline
\end{tabular}
\caption{\label{table-comparison}Comparison of some XLO and XFELO parameters at LCLS-II.}
\end{table}

Multi-bunch LCLS Cu linac capability has been demonstrated previously with up to 4 bunches and various time separations from 10 to 200 ns \cite{FJDecker, Decker:2015gry, Decker:2018oot}. In the low-gain regime, we expect to raise the number of linac pulses to 8.
The SASE pump creates a plasma channel 
of high temperature, giving rise to a shock wave (rarefaction wave)
and the lasing medium must be replenished before the next pump pulse arrives, and the stability of the jet has to be guaranteed. Recent experiments with slow water jets ($\leq$30 m/s) and LCLS Cu linac two-bunch mode \cite{Stan2016} revealed the shock wave travel time to be of the order of tens of nanoseconds. We are now testing similar effects with more powerful jets that have much higher speeds. The minimal replenishing time $T_{medium}$ will ultimately define the smallest realistic cavity size for XLO operation \cite{Wang:cn5057, Shapiro:pm5024}. 
Furhtermore, the cavity transit time, $T_{CAV}$, must also be a multiple of the linac RF period $T_{RF}$=350 ps: $\tau = T_{CAV}=nT_{RF}>T_{medium}$.  The advantages when using a small $\tau$ are  reduced time jitter in the SASE pump pulses, and more stability due to a compact cavity size. 
Finally, we provide a comparison of projected parameters of XLO operating at 8 keV and XFELO at 9.8 keV at LCLS-II \cite{Qin:FEL2017-TUC05, Kim:2019puj} in Tab. \ref{table-comparison}. Note that XFELO would be driven by a superconducting linac at a 1 MHz repetition rate and would have a new superconducting gun. XLO can operate on the room temperature linac at 120 Hz repetition rate. Its operation on the superconducting linac will be studied in another paper.

%\section{Summary}
In conclusion, we have studied how to realize an XLO operating in the 5 to 12 keV photon energy range using an XFEL pulse train as a periodic pump and a liquid jet gain medium. We discuss the oscillator's cavity, based on Bragg reflection optics in a bowtie configuration and explore the performance of a cupric nitrate lasing medium in various gain regimes. We show that XLO can generate intense, transversely coherent and nearly or fully transform limited X-ray pulses, with 48 meV FWHM bandwidth. The potential scientific applications of this novel X-ray source are extensive and include coherent imaging, inelastic scattering, X-ray interferometry.

\section{Acknowledgements}
Work was supported by the U.S. Department of Energy Contract No. DE-AC02-76SF00515. The authors wish to thank G. Marcus, A. Marinelli, Y. Shvyd'ko and Z. Huang for valuable suggestions throughout the research.

%\bibliographystyle{unsrt}
%\bibliography{biblio}

\begin{thebibliography}{100}

\bibitem{Hecht:08}
J. Hecht,
\newblock \href{http://www.osa-opn.org/abstract.cfm?URI=opn-19-5-26}{\em The History of the X-ray Laser}.
\newblock Opt. Photon. News, 5, pp. 26-33, 2008.

\bibitem{ChaplineWood}
G. Chapline, L. Wood,
\newblock \href{https://physicstoday.scitation.org/doi/10.1063/1.3069004}.
\newblock Physics Today, \textbf{28}, 6, p. 40, 1975.

\bibitem{PhysRevLett.54.110}
 D. L. Matthews, \textit{et al}, 
\newblock \href{https://link.aps.org/doi/10.1103/PhysRevLett.54.110}{Demonstration of a Soft X-Ray Amplifier}.
\newblock {\em Phys. Rev. Lett.}, 54, 2, pp. 110-113, 1985.

\bibitem{PhysRevLett.55.1753}
S. Suckewer,  C. H. Skinner, H. Milchberg, C. Keane, D. Voorhees,  
\newblock \href{https://link.aps.org/doi/10.1103/PhysRevLett.55.1753}{Amplification of stimulated soft x-ray emission in a confined plasma column}.
\newblock {\em Phys. Rev. Lett.}, 55, 17, pp. 1753-1756, 1985.

\bibitem{Suckewer_2009}
S. Suckewer and P. Jaegle,
\newblock Laser Physics Letters, \textbf{6}, 6, pp. 411-436, 2009.

\bibitem{PELLEGRINI1993223}
C. Pellegrini, \textit{et al},
\newblock \href{https://doi.org/10.1016/0168-9002(93)90047-L}{A 2 to 4 nm high power FEL on the SLAC linac}.
\newblock {\em NIM:A}, \textbf{331}, 1, pp. 223-227, 1993.

\bibitem{RevModPhys.88.015006}
C. Pellegrini, A. Marinelli, S. Reiche, 
\newblock \href{https://link.aps.org/doi/10.1103/RevModPhys.88.015006}{The physics of x-ray free-electron lasers}.
\newblock {\em Rev. Mod. Phys.}, \textbf{88}, 1, p. 015006, 2016.

\bibitem{RevModPhys.88.015007}
C. Bostedt, \textit{et al},
\href{https://link.aps.org/doi/10.1103/RevModPhys.88.015007}{Linac coherent light source: The first five years.}
\newblock {\em Rev. Mod. Phys.}, 88:015007, Mar 2016.

\bibitem{10.1039/9781782624097}
U. Bergmann, V. Yachandra, and J. Yano, editors.
\newblock \href{http://dx.doi.org/10.1039/9781782624097}{\em X-Ray Free Electron Lasers}.
\newblock Energy and Environment Series. The Royal Society of Chemistry, 2017.


\bibitem{Rohringer2012}
N. Rohringer, \textit{et al}
\newblock \href{https://www.nature.com/articles/nature10721}{Atomic inner-shell x-ray laser at 1.46 nanometres pumped by an x-ray  free-electron laser}.
\newblock {\em Nature}, 481:488 EP --, Jan 2012.

\bibitem{Yoneda}
H. Yoneda, \textit{et al},
\newblock \href{https://www.nature.com/articles/nature14894}{Atomic inner-shell laser at 1.5-Angström wavelength pumped by an x-ray free-electron laser}.
\newblock {\em Nature}, 524:446--449, 08 2015.

\bibitem{PhysRevLett.120.133203}
T. Kroll, \textit{et al},
\newblock \href{https://link.aps.org/doi/10.1103/PhysRevLett.120.133203}{Stimulated x-ray emission spectroscopy in transition metal complexes}.
\newblock {\em Phys. Rev. Lett.}, 120:133203, Mar 2018.

\bibitem{Kroll:16}
T. Kroll, \textit{et al},
\newblock \href{http://www.opticsexpress.org/abstract.cfm?URI=oe-24-20-22469}{X-ray absorption spectroscopy using a self-seeded soft x-ray free-electron laser}.
\newblock {\em Opt. Express}, 24(20):22469--22480, Oct 2016.

\bibitem{Bonifacio:1984qs}
R.~Bonifacio, C.~Pellegrini, and L.~Narducci.
\newblock\href{https://www.sciencedirect.com/science/article/pii/0030401884901056}{Collective Instabilities and High Gain Regime in a Free Electron  Laser}.
\newblock {\em Opt. Commun.}, 50:373--378, 1984.

\bibitem{DERBENEV1982415}
Ya.S. Derbenev, A.M. Kondratenko, and E.L. Saldin.
\newblock \href{http://www.sciencedirect.com/science/article/pii/0029554X82902336}{On the possibility of using a free electron laser for polarization of
  electrons in storage rings}.
\newblock {\em Nuclear Instruments and Methods in Physics Research}, 193(3):415-- 421, 1982.

\bibitem{PhysRevA.44.5178}
L.~H. Yu.
\newblock \href{https://link.aps.org/doi/10.1103/PhysRevA.44.5178}{Generation of intense UV radiation by subharmonically seeded single-pass free-electron lasers}.
\newblock {\em Phys. Rev. A}, 44:5178--5193, Oct 1991.

\bibitem{Allaria2012}
E. Allaria, \textit{et al},
\newblock \href{https://www.nature.com/articles/nphoton.2012.233?page=4}{Highly coherent and stable pulses from the fermi seeded free-electron laser in the extreme ultraviolet}.
\newblock {\em Nature Photonics}, 6:699–704, 09 2012.

\bibitem{Allaria2013}
E. Allaria, \textit{et al},
\newblock \href{https://www.nature.com/articles/nphoton.2013.277}{Two-stage seeded soft-x-ray free-electron laser}.
\newblock {\em Nature Photonics}, 7:913--918, 10 2013.

\bibitem{Stupakov2009}
G. Stupakov.
\newblock \href{https://journals.aps.org/prl/abstract/10.1103/PhysRevLett.102.074801}{Using the beam-echo effect for generation of short-wavelength  radiation}.
\newblock {\em Phys. Rev. Lett.}, 102 7:074801, 2009.

\bibitem{Ribic2019}
P. Ribic, \textit{et al},
\newblock \href{https://www.nature.com/articles/s41566-019-0427-1}{Coherent soft x-ray pulses from an echo-enabled harmonic generation free-electron laser}.
\newblock {\em Nature Photonics}, 13:1--7, 08 2019.

\bibitem{PhysRevLett.114.054801}
D.~Ratner,\textit{et al},
\newblock \href{https://link.aps.org/doi/10.1103/PhysRevLett.114.054801}{Experimental demonstration of a soft x-ray self-seeded free-electron laser}.
\newblock {\em Phys. Rev. Lett.}, 114:054801, Feb 2015.



\bibitem{doi:10.1080/09500340.2011.586473}
G. Geloni, V. Kocharyan, and E. Saldin.
\newblock \href{ https://doi.org/10.1080/09500340.2011.586473}{A novel self-seeding scheme for hard x-ray fels}.
\newblock {\em Journal of Modern Optics}, 58(16):1391--1403, 2011.

\bibitem{Amann}
J.~{Amann}, \textit{et al},
\newblock\href{https://www.nature.com/articles/nphoton.2012.180}{Demonstration of self-seeding in a hard-X-ray free-electron laser}.
\newblock {\em Nature Photonics}, 6:693--698, Oct 2012.

\bibitem{PhysRevLett.113.254801}
A.~A. Lutman, \textit{et al},
\newblock \href{https://link.aps.org/doi/10.1103/PhysRevLett.113.254801}{Demonstration of single-crystal self-seeded two-color x-ray free-electron lasers}.
\newblock {\em Phys. Rev. Lett.}, 113:254801, Dec 2014.

\bibitem{Osaka:rx5047}
T. Osaka, \textit{et al},
\newblock \href{https://doi.org/10.1107/S1600577519008841}{A micro channel-cut crystal X-ray monochromator for a self-seeded hard X-ray free-electron laser}.
\newblock {\em Journal of Synchrotron Radiation}, 26(5):1496--1502, Sep 2019.

\bibitem{Inoue2019}
I. Inoue, \textit{et al},
\newblock \href{https://www.nature.com/articles/s41566-019-0365-y}{Generation of narrow-band x-ray free-electron laser via reflection  self-seeding}.
\newblock {\em Nature Photonics}, 13:1, 05 2019.

\bibitem{PhysRevAccelBeams.22.080702}
G. Marcus, \textit{et al},
\newblock \href{https://link.aps.org/doi/10.1103/PhysRevAccelBeams.22.080702}{Experimental observations of seed growth and accompanying pedestal contamination in a self-seeded, soft x-ray free-electron laser}.
\newblock {\em Phys. Rev. Accel. Beams}, 22:080702, Aug 2019.

%\bibitem{PhysRevAccelBeams.23.010701}
%E. Hemsing, A. Halavanau, Z. Zhang,
%\newblock 
%\href{https://link.aps.org/doi/10.1103/PhysRevAccelBeams.23.010701}{Statistical theory of a self-seeded free electron laser with noise  pedestal growth.}
% \newblock {\em Phys. Rev. AB}, \textbf{23}, 1, p. 010701, 2020.


\bibitem{Colocho2013INCREASEDSR}
W.~Colocho, \textit{et al},
\href{http://accelconf.web.cern.ch/AccelConf/FEL2013/papers/wepso10.pdf}{Increased stability requirements for seeded beams at LCLS.} 
\newblock In Proc. of FEL13, New York City, NY, USA 2013.

\bibitem{REICHE200845}
S. Reiche, P. Musumeci, C. Pellegrini, J.B. Rosenzweig
\newblock 
\href{https://doi.org/10.1016/j.nima.2008.04.061}{Development of ultra-short pulse, single coherent spike for SASE X-ray FELs}
\newblock{\em NIM:A}, 593, 1, pp. 45-48, (2008).



\bibitem{PhysRevLett.92.074801}
P.~Emma, K.~Bane, M.~Cornacchia, Z.~Huang, H.~Schlarb, G.~Stupakov, and  D.~Walz.
\newblock \href{https://link.aps.org/doi/10.1103/PhysRevLett.92.074801}{Femtosecond and subfemtosecond x-ray pulses from a self-amplified spontaneous-emission--based free-electron laser}.
\newblock {\em Phys. Rev. Lett.}, 92:074801, Feb 2004.

\bibitem{doi:10.1063/1.4935429}
Y.~Ding, \textit{et al},
\newblock \href{https://doi.org/10.1063/1.4935429}{Generating femtosecond x-ray pulses using an emittance-spoiling foil in free-electron lasers}.
\newblock {\em Applied Physics Letters}, 107(19):191104, 2015.

\bibitem{doi:10.1063/1.4990716}
A.~Marinelli, J.~MacArthur, P.~Emma, M.~Guetg, C.~Field, D.~Kharakh, A.~A. Lutman, Y.~Ding, and Z.~Huang.
\newblock \href{https://doi.org/10.1063/1.4990716} {Experimental demonstration of a single-spike hard-x-ray free-electron  laser starting from noise}.
\newblock {\em Applied Physics Letters}, 111(15):151101, 2017.

\bibitem{Lutman2016}
A.~A. Lutman, \textit{et al}, 
\newblock \href{https://www.nature.com/articles/nphoton.2016.201}{Fresh-slice multicolour x-ray free-electron lasers}.
\newblock {\em Nature Photonics}, 10:745 EP --, Oct 2016.
\newblock Article.

\bibitem{Lutman:2018cot}
A.~A. Lutman, \textit{et al}, 
\newblock \href{https://journals.aps.org/prl/abstract/10.1103/PhysRevLett.120.264801}{High-Power Femtosecond Soft X Rays from Fresh-Slice Multistage  Free-Electron Lasers}.
\newblock {\em Phys. Rev. Lett.}, 120(26):264801, 2018.

\bibitem{PhysRevLett.119.154801}
S.~Huang, \textit{et al}, 
\newblock \href{https://link.aps.org/doi/10.1103/PhysRevLett.119.154801}{Generating single-spike hard x-ray pulses with nonlinear bunch compression in free-electron lasers}.
\newblock {\em Phys. Rev. Lett.}, 119:154801, Oct 2017.

\bibitem{JoeNature}
J. Duris, \textit{et al}, 
\href{https://doi.org/10.1038/s41566-019-0549-5}{Tunable isolated attosecond X-ray pulses with gigawatt peak power from a free-electron laser.}
\newblock {\em Nature Photonics}, 1749-4893, 2019.

\bibitem{PhysRevAccelBeams.20.030701}
C. Emma, Y. Feng, D. C. Nguyen, A. Ratti, C. Pellegrini.
\newblock \href{https://link.aps.org/doi/10.1103/PhysRevAccelBeams.20.030701}{Compact double-bunch x-ray free electron lasers for fresh bunch self-seeding and harmonic lasing}.
\newblock {\em Phys. Rev. Accel. Beams}, \textbf{20}, 3, p. 030701, 2017

\bibitem{doi:10.1063/1.4980092}
C. Emma, A. A. Lutman, M. Guetg, J. Krzywinski, A. Marinelli, J. Wu, C Pellegrini.
\newblock \href{https://doi.org/10.1063/1.4980092}{Experimental demonstration of fresh bunch self-seeding in an X-ray free electron laser}.
\newblock {\em Applied Physics Letters}, \textbf{110}, 15, p. 154101, 2017 

\bibitem{Halavanau:co5121}
A. Halavanau, F.-J. Decker, C. Emma, J. Sheppard, and
  C. Pellegrini.
\newblock \href{https://doi.org/10.1107/S1600577519002492}{Very high brightness and power LCLS-II hard X-ray pulses}.
\newblock {\em Journal of Synchrotron Radiation}, 26(3), May 2019.

\bibitem{PhysRevLett.100.244802}
K.-J. Kim, Y. Shvyd'ko, and S. Reiche.
\newblock \href{https://link.aps.org/doi/10.1103/PhysRevLett.100.244802}{A proposal for an x-ray free-electron laser oscillator with an  energy-recovery linac}.
\newblock {\em Phys. Rev. Lett.}, 100:244802, Jun 2008.

\bibitem{COLELLA198441}
R.~Colella and A.~Luccio.
\newblock \href{http://www.sciencedirect.com/science/article/pii/0030401884900099}{Proposal for a free electron laser in the x-ray region}.
\newblock {\em Optics Communications}, 50(1):41 -- 44, 1984.

\bibitem{PhysRevLett.96.144801}
Z. Huang and R.~D. Ruth.
\newblock \href{https://link.aps.org/doi/10.1103/PhysRevLett.96.144801}{Fully coherent x-ray pulses from a regenerative-amplifier free-electron laser}.
\newblock {\em Phys. Rev. Lett.}, 96:144801, Apr 2006.

\bibitem{Adams:2019njp}
B. Adams, \textit{et al},
\newblock \href{https://arxiv.org/abs/1903.09317}{Scientific Opportunities with an X-ray Free-Electron Laser Oscillator}.
\newblock arXiv:1903.09317, 2019.

\bibitem{XOPcode}
M.~S. del Rio and R.~J. Dejus.
\newblock\href{https://doi.org/10.1117/12.893911}{Xop v2.4: recent developments of the x-ray optics software toolkit}.
\newblock {\em Proc.SPIE}, 8141:8141 -- 8141 -- 5, 2011.

\bibitem{UweKB}
T. Kroll, \textit{et al},
%Uwe's Kbeta experiment
\newblock submitted to {\em PRL}.

\bibitem{PhysRevA.99.013839}
A. Benediktovitch, V.~P. Majety, and N. Rohringer.
\newblock\href{https://journals.aps.org/pra/abstract/10.1103/PhysRevA.99.013839}{Quantum theory of superfluorescence based on two-point correlation functions}.
\newblock {\em Phys. Rev. A}, 99:013839, Jan 2019.

\bibitem{PhysRevLett.123.023201}
L.~Mercadier, \textit{et al},
\newblock \href{https://link.aps.org/doi/10.1103/PhysRevLett.123.023201}{Evidence of extreme ultraviolet superfluorescence in Xenon}.
\newblock {\em Phys. Rev. Lett.}, 123:023201, Jul 2019.

\bibitem{Jurek:zd5003}
Z. Jurek, S.-K. Son, B.Ziaja, and R. Santra.
\newblock \href{https://doi.org/10.1107/S1600576716006014}{{\it XMDYN} and {\it XATOM}: versatile simulation tools for quantitative modeling of X-ray free-electron laser induced dynamics of matter}.
\newblock {\em Journal of Applied Crystallography}, 49(3):1048--1056, Jun 2016.

\bibitem{PhysRevA.83.033402}
S.-K. Son, L. Young, and R. Santra.
\newblock \href{https://link.aps.org/doi/10.1103/PhysRevA.83.033402}{Impact of hollow-atom formation on coherent x-ray scattering at high intensity}.
\newblock {\em Phys. Rev. A}, 83:033402, Mar 2011.

\bibitem{RevModPhys.39.125}
J.~A. Bearden and A.~F. Burr.
\newblock \href{https://link.aps.org/doi/10.1103/RevModPhys.39.125}{Reevaluation of x-ray atomic energy levels}.
\newblock {\em Rev. Mod. Phys.}, 39:125--142, Jan 1967.

\bibitem{GLATZEL200565}
P. Glatzel and U. Bergmann.
\newblock \href{http://www.sciencedirect.com/science/article/pii/S0010854504001146}{High resolution 1s core hole x-ray spectroscopy in 3d transition metal complexes—electronic and structural information}.
\newblock {\em Coordination Chemistry Reviews}, 249(1):65 -- 95, 2005.
\newblock Synchrotron Radiation in Inorganic and Bioinorganic Chemistry.

\bibitem{doi:10.1063/1.1651874}
R.~M.~J. Cotterill.
\newblock \href{https://doi.org/10.1063/1.1651874}{A universal planar x‐ray resonator}.
\newblock {\em Applied Physics Letters}, 12(12):403--404, 1968.

\bibitem{Kolodziej:te5013}
T. Kolodziej, P. Vodnala, S. Terentyev, V. Blank, and Y.
  Shvyd'ko.
\newblock \href{https://doi.org/10.1107/S1600576716009171}{Diamond drumhead crystals for X-ray optics applications}.
\newblock {\em Journal of Applied Crystallography}, 49(4):1240--1244, Aug 2016.

\bibitem{Shvydko:2019gvv}
Y. Shvyd'ko.
\newblock \href{https://link.aps.org/doi/10.1103/PhysRevAccelBeams.22.100703}{Output coupling from x-ray free-electron laser cavities with  intracavity beam splitters}.
\newblock {\em Phys. Rev. Accel. Beams}, 22:100703, Oct 2019.

\bibitem{Freund_2019}
H. P. Freund and P. J. M. van der Slot and Yu. Shvyd'ko
\newblock {An x-ray regenerative amplifier free-electron laser using diamond pinhole mirrors}.
\newblock {\em New Journal of Physics}, \textbf{21}, 9, p. 093028, Oct 2019.

\bibitem{RevModPhys.36.681}
B.~W. Batterman and H. Cole.
\newblock \href{https://journals.aps.org/rmp/pdf/10.1103/RevModPhys.36.681}{Dynamical diffraction of X-rays by perfect crystals}.
\newblock {\em Rev. Mod. Phys.}, 36:681--717, Jul 1964.

\bibitem{PhysRevSTAB.15.050706}
R.~R. Lindberg and Yu.~V. Shvyd'ko.
\newblock \href{https://journals.aps.org/prab/abstract/10.1103/PhysRevSTAB.15.050706}{Time dependence of bragg forward scattering and self-seeding of hard
  x-ray free-electron lasers}.
\newblock {\em Phys. Rev. ST Accel. Beams}, 15:050706, May 2012.

\bibitem{PhysRevSTAB.15.100702}
Y. Shvyd'ko and R. Lindberg.
\newblock \href{https://link.aps.org/doi/10.1103/PhysRevSTAB.15.100702}{Spatiotemporal response of crystals in x-ray bragg diffraction}.
\newblock {\em Phys. Rev. ST Accel. Beams}, 15:100702, Oct 2012.

\bibitem{PhysRev.137.A1869}
J.~J. DeMarco and R.~J. Weiss.
\newblock \href{https://link.aps.org/doi/10.1103/PhysRev.137.A1869}{Absolute x-ray scattering factors of silicon and germanium}.
\newblock {\em Phys. Rev.}, 137:A1869--A1871, Mar 1965.

%\bibitem{PATIMISCO201770}
%P. Patimisco, A. Sampaolo, F.~K. Tittel, and V. Spagnolo.
%\newblock \href{https://doi.org/10.1016/j.sna.2017.10.005}{Mode matching of a laser-beam to a compact high finesse bow-tie optical cavity for quartz enhanced photoacoustic gas sensing}.
%\newblock {\em Sensors and Actuators A: Physical}, 267:70 -- 75, 2017.

\bibitem{1071268}
G.~{Dattoli}, A.~{Marino}, A.~{Renieri}, and F.~{Romanelli}.
\newblock \href{https://ieeexplore.ieee.org/document/1071268}{Progress in the hamiltonian picture of the free-electron laser}.
\newblock {\em IEEE Journal of Quantum Electronics}, 17(8):1371--1387, 1981.

\bibitem{1072746}
P.~{Elleaume}.
\newblock \href{https://ieeexplore.ieee.org/document/1072746} {Microtemporal and spectral structure of storage ring free-electron lasers}.
\newblock {\em IEEE Journal of Quantum Electronics}, 21(7):1012--1022, July 1985.

\bibitem{FJDecker}
F.~J. Decker, \textit{et al},
\newblock \href{https://accelconf.web.cern.ch/AccelConf/FEL2010/papers/wepb33.pdf}{A demonstration of multi-bunch operation in the LCLS}.
\newblock In {\em {Proceedings of FEL2010, Malmoe, Sweden}}, page 467, 2010.

\bibitem{Decker:2015gry}
F.-J. Decker, \textit{et al},
\newblock \href{http://accelconf.web.cern.ch/AccelConf/FEL2015/papers/wep023.pdf}{Two Bunches with ns-Separation with LCLS}.
\newblock In {\em {Proceedings, 37th International Free Electron Laser  Conference, Daejeon, Korea, August 23-28}}, page WEP023,
  2015.

\bibitem{Decker:2018oot}
F.-J. Decker, K. Bane, W. Colocho, A. A. Lutman, and J.
  Sheppard.
\newblock \href{http://inspirehep.net/record/1671673/files/tup023.pdf}{Recent Developments and Plans for Two Bunch Operation with up to 1  $\mu$s Separation at LCLS}.
\newblock In {\em {Proceedings, 38th International Free Electron Laser Conference, FEL2017}}, page TUP023, 2018.

\bibitem{Stan2016}
C.~A. Stan, \textit{et al},
\newblock \href{https://www.nature.com/articles/nphys3779}{Liquid explosions induced by x-ray laser pulses}.
\newblock {\em Nature Physics}, 12:966 EP, 2016.

\bibitem{Wang:cn5057}
D. Wang, U. Weierstall, L. Pollack, Lois and J. Spence
\newblock \href{https://doi.org/10.1107/S160057751401858X}{Double-focusing mixing jet for XFEL study of chemical kinetics}.
\newblock {\em J. of Synchr. Rad.}, 21, 6, pp. 1364-1366, 2014.

\bibitem{Shapiro:pm5024}
D. A. Shapiro, \textit{et al},
\newblock \href{https://doi.org/10.1107/S0909049508024151}{Powder diffraction from a continuous microjet of submicrometer protein crystals}.
\newblock {\em J. of Synchr. Rad.}, 15, 6, pp. 593-599, 2008.

\bibitem{Qin:FEL2017-TUC05}
   W. Qin, K.L.F. Bane, Y. Ding, S. Huang, Z. Huang, K.-J. Kim, \emph{et al.},
%   W. Qin, K.L.F. Bane, Y. Ding, S. Huang, Z. Huang, K.-J. Kim, R.R. Lindberg, K.X. Liu, G. Marcus, and T.J. Maxwell,
%   W. Qin \emph{et al.},
   \textquotedblleft{Start-to-End Simulations for an X-Ray FEL Oscillator at the LCLS-II and LCLS-II-HE}\textquotedblright,
   in \emph{Proc. FEL2017}, \url{https://doi.org/10.18429/JACoW-FEL2017-TUC05}, 2018.

\bibitem{Kim:2019puj}
K. J. Kim, \textit{et al},
\newblock \href{http://accelconf.web.cern.ch/AccelConf/ipac2019/papers/tuprb096.pdf}{Test of an X-ray Cavity using Double-Bunches from the LCLS Cu-Linac}.
\newblock {\em Proc. of IPAC19}, paper TUPRB096, Melbourne, Australia, May 19-24, 2019.


\end{thebibliography}

\end{document}